\documentclass[twocolumn,preprintnumbers,superscriptaddress,showpacs,showkeys,nofootinbib]{revtex4}
\usepackage{amsmath}
\usepackage{bm}
\usepackage{amssymb}
\usepackage{graphicx}
\usepackage{hyperref}
\usepackage{xspace}
\usepackage{bm}
\usepackage{slashed}
\usepackage{xspace}
\usepackage{color}

\usepackage{booktabs}
\AtBeginDocument{
\heavyrulewidth=.08em
\lightrulewidth=.05em
\cmidrulewidth=.03em
\belowrulesep=.65ex
\belowbottomsep=0pt
\aboverulesep=.4ex
\abovetopsep=0pt
\cmidrulesep=\doublerulesep
\cmidrulekern=.5em
\defaultaddspace=.5em
}

\newcommand{\be}{\begin{equation}}
\newcommand{\ee}{\end{equation}}
\newcommand{\ba}{\begin{eqnarray}}
\newcommand{\ea}{\end{eqnarray}}

\usepackage{xcolor}
\definecolor{light-gray}{gray}{0.8}

\newcommand{\bbb}[1]{\boldsymbol{#1}}

\begin{document}

\title{$B_c^\pm$ decays into tetraquarks}
 
\preprint{DESY 16-045/ April 2016}

\newcommand{\sapienza}{Dipartimento di Fisica, `Sapienza' Universit\`a di Roma \\
P.le Aldo Moro 5, I-00185 Roma, Italy}
\newcommand{\cern}{CERN, Theory Division, Geneva 23, Switzerland}
\newcommand{\desy}{Deutsches Elektronen-Synchrotron DESY, D-22607 Hamburg, Germany}
\newcommand{\infn}{INFN Sezione di Roma, P.le Aldo Moro 5, I-00185 Roma, Italy}

\author{A.~Ali}
\affiliation{\desy}
\author{L.~Maiani}
\affiliation{\sapienza}
\affiliation{\infn}
\author{A.D.~Polosa}
\affiliation{\sapienza}
\affiliation{\infn}
\affiliation{\cern}
\author{V.~Riquer}
\affiliation{\sapienza}
\affiliation{\infn}

\begin{abstract}
The recent observation by the D0 collaboration of a narrow structure $X(5568)$
consisting of four different quark flavors $bdus$, has not been confirmed by LHCb. More data and dedicated
 analyses are needed to cover a larger mass range.
 In the tightly bound diquark model, we estimate the lightest $bdus$, $0^+$ tetraquark 
 at  a mass of about 5770~MeV, 
approximately 200~MeV above the reported $X(5568)$, and just 7~MeV below the $B \bar{K}$
threshold. The charged tetraquark is accompanied by $I=1$ and $I=0$ neutral partners almost degenerate in mass. A $bdus$, $S$-wave, $1^+$ quartet at $5820$~MeV is implied as well. In the charm sector, $cdus$, $0^+$ and $1^+$ tetraquarks are predicted at 
$2365$~MeV and $2501$~MeV,  about $40-50$~MeV heavier than $D_{s0}(2317)$ and $D_{s1}(2460)$.
$bdus$ tetraquarks can be searched  in the hadronic debris of a jet initiated by a $ b $. However,
some of them may also be produced in $B_c$ decays. The proposed 
discovery modes of $S$-wave tetraquarks are 
$B_c \to X_{b0} + \pi$ 
 with the subsequent decays 
 $X_{b0} \to B_s + \pi$, giving rise to final states such as $B_s \pi^+ \pi^0$. 
 We also emphasize the importance of $B_c$ decays as a source of bound hidden charm tetraquarks, 
 such as $B_c \to X(3872) + \pi$.
 \end{abstract}

\maketitle

\section{Introduction} \label{intro}
Recently, the D0 experiment reported the observation of a new narrow structure in the $B_s^0\,\pi^+$ invariant mass\footnote{Hereafter, adding the charged conjugated modes --- \emph{e.g.}
 ${\bar B}_s^0\pi^-$ --- is understood.}~\cite{D0:2016mwd}, which promptly attracted considerable
 attention, see~\cite{others} (but skepticism has been raised in \cite{Burns:2016gvy}). Based on 
 10.4 fb$^{-1}$ of $p\bar{p}$ collision data at $\sqrt{s}=1.96$ TeV,
 this candidate resonance, dubbed $X(5568)$,
 has a mass and width given by $M=5568$~MeV and $\Gamma=22$~MeV, respectively.

 A state  such as $X(5568)$ would be distinct in that a charged light quark pair cannot be created from the vacuum, leading to the unambiguous composition in terms of  four valence quarks with different flavors --- $\bar b\bar d s u$ (tetraquarks with flavored quantum numbers have also been discussed in~\cite{tantaloni}).
 
Exciting a discovery as it would have been, $X(5568)$ has not been confirmed by the LHCb experiment. Their analysis has been reported  recently, based on 3 fb$^{-1}$ of $pp$ collision data
at $\sqrt{s}=7$ and 8 TeV, yielding a data sample of $B_s^0$ mesons 20 times higher than that of the D0 collaboration.
 Adding then a charged pion, the $B_s^0 \pi^+$ invariant mass shows no
structure from the $B_s^0 \pi^+$ threshold up to
 $M_{B_s^0 \pi^+} \leq 5700$~MeV~and an upper limit on the ratio
  $\rho(X(5568)/B_s^0) < 0.016 (0.018)$ @ 90 (95) \% C.L. is set for $p_T(B_s^0) > 10$ GeV
~\cite{LHCb:2016ppf}. 

The valence quark composition of $X(5568)$ fits into a diquarkonium interpretation~\cite{tetra0,tetra1,tetra2,tetrab,tetra3}. In this framework, the constituents are arranged in a tightly bound diquark-antidiquark pair, $[ \bar b \bar d]_{\bbb{3}_c}[s u]_{\bar{\bbb{3}}_c} $, both of them transforming non-trivially under color SU(3). The possible manifestation of these compact tetraquarks follows essentially from symmetry considerations as in the original constituent quark model and their spectrum is rich. However, as outlined below, our computation of the tetraquark mass spectrum with
 the quark flavors $\bar b\bar d s u$ yields
significantly higher values. The lightest in this sector is the $S$-state, $X_{b0}^+$, whose mass is 
estimated by us to be about 5770~MeV - approximately
200~MeV heavier than the X(5568), and below the $B^+ \bar{K}^0$ threshold by about 7~MeV.

The tetraquark mass spectrum is calculable up to a theoretical error which we estimate to be of the order of $\pm ~30$~MeV, judging from the discrepancies of constituent quark masses obtained from baryons and mesons (see e.g. Table I in ref.~\cite{tetra1}). Thus, $X_{b0}^+$ and $X_{b0}^0$ may lie somewhat above the $B^+\bar{K}^0$ threshold, in which case $X_{b0}^+$  will decay, perhaps mostly, in the   $B^+ \bar{K}^0$ mode, and
the $B_s^0 \pi^+$ resonance signal would be reduced\footnote{This would be similar to the case of $X(3278)$, which decays predominantly in $D D^*$ and also, appreciably, in $J/\psi+\rho/\omega$.}.  An analysis of the $B^+K^-$ final state has been published by LHCb, based on a limited sample of 1~fb$^{-1}$~\cite{lhcbat}. 

However, it is also within the margin of errors  that the actual masses of these tetraquark $S$-states are couple of tens of MeV  below our estimates, in which case, the $B^+ \bar{K}^0$ mode is not available, and it is logical to anticipate $X_{b0}^+$ and $X_{b0}^0$ as resonant $B_s \pi$ states.
We pursue this possibility here. An alternative description is found in~\cite{fm}.

\begin{figure*}[ht!]
\centering
\includegraphics[width=1.1\columnwidth]{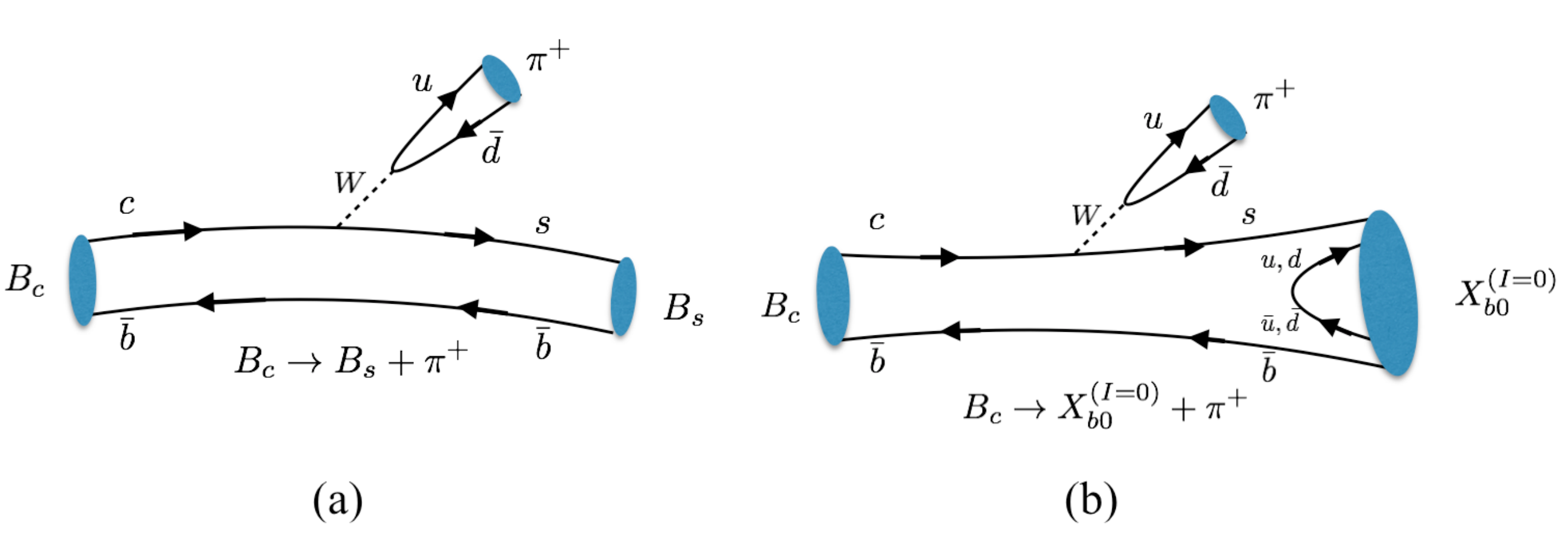}
\includegraphics[width=0.8\columnwidth]{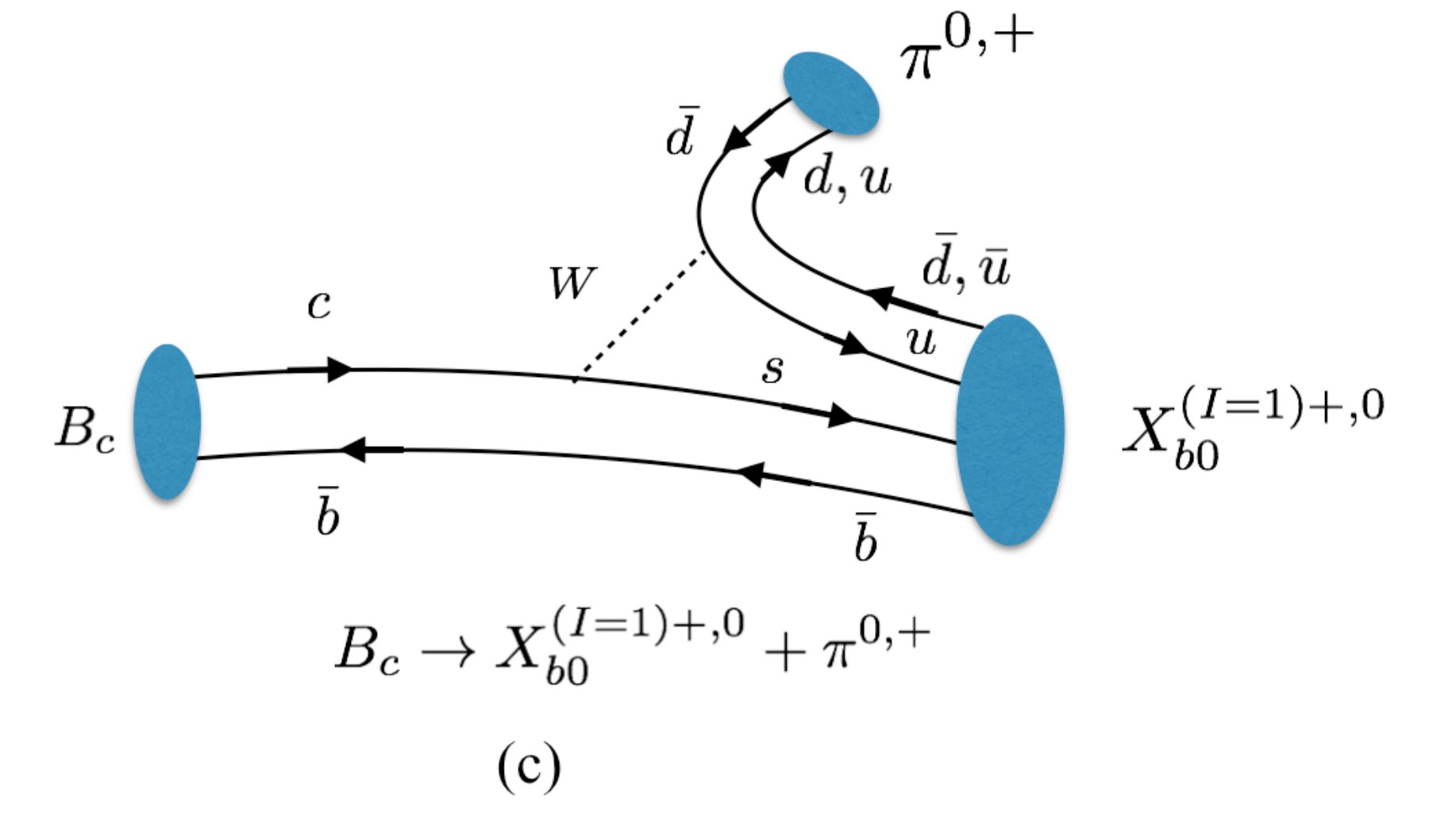}
\caption{(a): Leading order Feynman diagram for the decay $B_c^+ \to B_s^0 \pi^+$  (b):
 $B_c^+ \to X_{b0}^{(I=0)} + \pi^+$
(c): the corresponding diagram for the decays $B_c^+ \to X_{b0}(5570)^{(I=1)+,0} + \pi^{0,+}$. 
} 
\label{Feyn}
\end{figure*}

 With this hindsight,
we point out that there are, in principle, two {\it generic different mechanisms} for 
producing $X_{b0}(5770)$  in high energy $pp$ and $p\bar{p}$ collisions.
These states can be produced as a fragmentation product of a jet initiated by a
$b$-quark, but, subject to phase space,  they can also be produced in the decays of the $B_c^\pm$ mesons, 
$B_c^\pm \to X_{b0}(5770)^{I=1} + \pi$ and $B_c^\pm \to X_{b0}(5770)^{I=0} + \pi^\pm$
as a result of weak ($c \to s u \bar{d})$ decays, $q\bar{q}$ excitation, and quark rearrangement
(see Fig.~\ref{Feyn}).
With the anticipated decays 
\mbox{$X_{b0}^\pm \to B_s^0 \pi^\pm$} and \mbox{$X_{b0}^0 \to B_s^0 \pi^0$},
the decay chains will lead to \mbox{$B_c^\pm \to B_s^0 \,\pi^\pm \pi^0$} etc.
 A resonating structure in the $B_s \pi$ mode can then be fished out
by Dalitz analysis.
 This mechanism is similar to 
the production mechanism of many multiquark states, seen in $B^0$ and $B^\pm$ decays, such as
 $B \to X(3872) (K, K \pi)$, but also for the pentaquarks, such as $P_c(4450)^+$
and $P_c(4380)^+$, in the decays $\Lambda_b^0 \to (P_c(4380)^+, P_c(4450)^+) K^-$.
We recall that the dominant two-body decay mode
$B_c^\pm \to B_s^0 \pi^\pm$ has been measured
by LHCb, with a branching ratio of about 10\%~\cite{Aaij:2013cda}, and we anticipate that
some of the $B_c^\pm$-decays to tetraquarks will be large enough to be measured. 
 
In what follows, we present our estimates of the mass spectrum  of the lowest $S$ and $P$-states
 with the flavor quantum numbers of
the state  $B_s^0 \pi^+ = (\bar{b}s)(\bar{d}u$), having the angular momentum quantum
 numbers $J^P=0^+, 1^+$ together with their counterparts in the charm sector.
 This is followed by the discussion of the $B_c^\pm $-decays leading to some of
 these tetraquark states as well as the bound $c\bar{c}$ tetraquark states $X(3872)$
 in the decays $B_c^\pm \to X(3872) + \pi^\pm$.

\section{Spectrum} \label{spectr}
Within the constituent quark model the color-spin Hamiltonian describing the interaction between the different constituents of a hadron takes the form
\begin{align} \label{H}
H=\sum_i m_i+2\sum_{i<j}\kappa_{ij}\,\bm{S}_i\cdot \bm{S}_j
\end{align}
where $m_i$ are the diquark constituent masses, $\bbb{S}_i$ the quark spins and $\kappa_{ij}$ some effective, representation-dependent chromomagnetic couplings. The spin-spin interaction is here understood to be a contact one.

In the most recent and most successful type-II tetraquark model~\cite{tetra2,tetra3}, the dominant interactions are assumed to be the spin-spin interactions between quarks (antiquarks) inside the same tightly bound diquark (antidiquark). With the composition:  $[\bar b\bar q]_{{\bbb{3}}_c}[ s q^\prime]_{\bar{\bbb{3}}_c}$with \mbox{$q\neq q^\prime =d,u $}, this means retaining only $\kappa_{bq}$ and $\kappa_{sq^\prime}$ and the lightest states will correspond to the heavy-light diquark spins: $S_{[bq]}=0,1$ and $S_{[sq]}=0$. The latter case corresponds to the so called `good diquark'~\cite{jaffe}, and the two resulting states have $J^P=0^+$ or $1^+$, the lightest  being the $0^+$ one. To indicate these particles, we use the notations
\be
X_{b0}=|0_{\bar b \bar q}, 0_{sq^\prime}\rangle\quad\quad X_{b1}=|1_{\bar b \bar q}, 0_{sq^\prime}\rangle
\label{notation}
\ee

In the above approximation, the resulting mass formula for $S$-wave, $[\bar b\bar q][s q^\prime]$ states is {\it additive in diquark energies}, 
\ba 
&& M(X_{bS})= m_{[bq]} +2\kappa_{bq}\,\bm{S}_{\bar b}\cdot \bm{S}_{\bar q}+m_{[sq]} +2\kappa_{sq}\,\bm{S}_s\cdot \bm{S}_{q^\prime}\nonumber \\
&&=m_{[bq]} +\kappa_{bq}\left(S(S+1)-{\footnotesize \frac{3}{2}}\right)+m_{[sq]} -{\footnotesize  \frac{3}{2}} \kappa_{sq} 
\label{bs}
\ea
where $S\equiv S_{[bq]}$.

We may compare (\ref{bs}) with the mass formulae of the related tetraquarks $a_0(980)$\cite{tetra0}, $Z_b(10610), Z^\prime_b (10650)$\cite{tetrab}, obtained with the substitutions: $b\bar s \to s \bar s$ and $b\bar s \to b \bar b$
\ba 
&& a_0(980)=|0_{\bar s \bar q}, 0_{sq^\prime}\rangle\label{ascal}\\
&& \quad M_{a_0}=2\left(m_{[sq]} -\frac{3}{2}\, \kappa_{sq}\right)\notag  \\
&&\notag\\
&&Z_b=\frac{1}{\sqrt{2}}\left( |1_{\bar b \bar q}, 0_{bq^\prime}\rangle-|0_{\bar b \bar q}, 1_{bq^\prime}\rangle\right)\label{zetab} \\
&&\quad M_{Z_b}=2\,m_{[bq]}-\kappa_{bq} \notag\\
&&\notag\\
&&Z^\prime_b= |1_{\bar b \bar q}, 1_{bq^\prime}\rangle_{J=1}\label{zetabp}\\
&& \quad M_{Z^\prime_b}=2\,m_{[bq]}+\kappa_{bq}\notag
\ea

From Eqs.~(\ref{zetab}) and (\ref{zetabp}) and the known masses~\cite{pdg}, we derive 
\begin{subequations}
\begin{align}
m_{[bq]}&=\frac{M(Z_b^\prime)+M(Z_b)}{4}\simeq 5315~\text{MeV} \\
\kappa_{bq}&=\frac{M(Z_b^\prime)-M(Z_b)}{2}\simeq 22.5~\text{MeV}\label{kappabq}
\end{align}
\end{subequations}

In the approximation where tetrquark masses are additive in diquark energies, one finds
\ba
M(X_{b0})&=&\left(m_{[bq]} -{\footnotesize  \frac{3}{2}} \kappa_{bq}\right)_{Z_b}+\left(m_{[sq]} -{\footnotesize  \frac{3}{2}} \kappa_{sq}\right)_{a_0}=\nonumber\\
& \simeq& 5770~ {\rm MeV}~(J^P=0^+)
\label{xbo}
\ea
about $200$~MeV more than the X(5568) mass and just 7~MeV below the $B^+ \bar{K}^0$. 

To be seen as resonant $B_s\pi$ states, their masses should lie below the $BK$ threshold. A good part of the $B_s\pi$ invariant mass spectrum is excluded by the LHCb, but still there is a window of opportunity left unexplored so far.

As a side remark, we note that in Ref.~\cite{tetra1} the value $m_{[sq]}=590$~MeV was obtained using the value  \mbox{$\kappa_{sq}\simeq 64$~MeV} obtained from a fit to the baryon masses, which however may be different from the spin-spin coupling inside a diquark. On the other hand,  $\kappa_{ij}$ are expected to scale inversely to  the constituent quark  masses and this relation is approximately verified by $\kappa_{bq}$ and $\kappa_{cq}$~\cite{tetra3} estimated from $Z_{b,c}$ and $Z^\prime_{b,c}$ masses, eq.~(\ref{kappabq}) and eq.~(\ref{kappacq}) below. If we scale $\kappa_{sq}$ from $\kappa_{cq}$ using the strange and charm constituent quark masses, we obtain
\be
\kappa_{sq}\simeq 200~{\rm MeV}\quad 
\ee
leading~to
\be 
\quad m_{[sq]}\simeq 800~{\rm MeV}
\ee
The diquark mass thus obtained is close to the sum of constituent light and strange quark masses, $330$ and $520$~MeV, respectively.

The $J^P=1^+$ exotic states lies close by. From Eq.~(\ref{bs}) we find
\be
 M (X_{b1})  \simeq 5820~{\rm MeV}~(J^P=1^+)
 \ee
 
The $X_{b1}$ state is expected to decay into $B_s^{*0}\pi^+$ followed by $B_s^{*0}\to B_s^0\gamma$, with a photon energy of 48~MeV in the $B_s^*$ rest frame. Such a low energy photon escapes detection at hadron colliders, as pointed out in~\cite{D0:2016mwd}. As a consequence of this,
 the observed peak of the $X_{b1}$ would be shifted towards lower invariant masses and essentially coincide with the $X_{b0}$ peak.

In the type-II model~\cite{tetra2}, we estimate the parameters $m_{[cq]}$ and $\kappa_{cq}$,  from the masses of $Z_c(3900)$, $Z_c^\prime(4020)$~\cite{pdg}, obtaining
\begin{subequations}
\begin{align}
m_{[cq]}&=\frac{M(Z_c^\prime)+M(Z_c)}{4}\simeq 1978~\text{ MeV} \\
\kappa_{cq}&=\frac{M(Z_c^\prime)-M(Z_c)}{2}\simeq 67~\text{ MeV}\label{kappacq}
\end{align}
\end{subequations}

One might use the previous results to estimate the mass  of the analogous $X_{cS}^\pm$ expected in the charm sector and decaying into $D_s\,\pi$
\ba
M(X_{c0}) &=&  m_{[cu]}+ m_{[sd]}-3/2\, \kappa_{sq} -3/2\, \kappa_{cq}  \nonumber \\
&\simeq&~ 2367\,\text{MeV}
\label{c0prediction}\\
M(X_{c1})&=& m_{[cu]}+ m_{[sd]}-3/2\, \kappa_{sq} +1/2\, \kappa_{cq}  \nonumber \\
&\simeq&~2501\,\text{MeV}
\label{c1prediction}
\ea

The estimates in Eq.~(\ref{c0prediction} -\ref{c1prediction}) set the exotic candidates $X_{c0}^\pm$ just above the $D\,K$  and $D^*\,K$ thresholds ($2363$ and $2504$~MeV, respectively), so that it could be useful to search also in these decay channels. 

If the light diquark is in the $S=0$ configuration, {\it i.e.} it is antisymmetric in spin and color, it must also be antisymmetric in SU(3)$_{\mathrm{F}}$ (F for flavor), therefore the tetraquarks $[\bar Q\bar q][q^\prime q^{\prime\prime}]$, with $Q=b, c$ and $q,q^\prime, q^{\prime\prime}=u,d,s$ belong to the  SU(3)$_{\mathrm{F}}$ representation:  $ \bar {\bm 3}\otimes\bar {\bm  3}= {\bm 3}\oplus  \bar {\bm 6}$.

In the charm sector, one doubly charged state is present, belonging to the $\bar {\bm 6}$, {\it e.g.} with the flavor content $[\bar c \bar u][sd]\to D_s^-\pi^-$. 
In the beauty sector, doubly charged states lie in the symmetric {$\bm{15}$} representation of  $SU(3)_{\mathrm{F}}$ (see,  He and Ko in~\cite{others}), originating from the product: $ \bar {\bm 3}\otimes{\bm  6}= {\bm 3}\oplus  {\bm{15}}$.
 This requires a light diquark with $S=1$, the so-called ``bad diquarks'', which may be argued to have little binding~\cite{jaffe}. 
 
At present, upper limits on the production at lepton colliders of charmed-strange doubly charged resonances have been given~\cite{lepton} in the $D_s^+\pi^+$ channel,  for masses between 2.25 and 2.61 GeV.

We close this Section by considering the flavour multiplicity of the states $X_{b0}=[\bar b \bar q][s q^\prime]$, with $q, q^\prime=u,d$, and their decay modes. These  states are obviously organised in a isospin triplet and singlet, similar in structure to the scalar light tetraquarks $a_0(980)$ and $f_0(980)$. The neutral $X_{b0}$ states are similarly expected to be nearly degenerate in mass. 

The isoscalar state should decay as $X_{b0}^{(I=0)} \to B_s +\eta$ which is most likely  phase space forbidden, leaving the possibility of the strong decay  $X_{b0}^{(I=0)} \to B+\bar K$, a situation very similar to the decay $f_0 \to K \bar K$. Should also the latter mode be forbidden by phase space, $X_{b0}^{(I=0)}$  has to decay by isospin violating interactions: \mbox{$X_{b0}^{(I=0)}\to B_s +\pi^0$}, which may occur due to isospin violating mixing with $X_{b0}^{(I=1)}$ or via $\eta-\pi^0$ mixing, similarly to $\eta$ decay.

Similar considerations apply to the $I=0$ $X_{cS}^\pm$ states which estimates in Eqs.~(\ref{c0prediction} -\ref{c1prediction}) place only $40-50$~MeV above the well known $D_{s0}(2317)$ and $D_{s1}(2460)$. The mass difference is quite close to the theoretical error so as to suggest $X_{cS}$ to be identified with the latter resonances, the decays into $D_s^+ \pi^0$ arising also from isospin breaking interactions, either due to the mixing with the $I=1, I_3=0$ component or via $\eta-\pi^0$ mixing.

\section{Tetraquark production in weak decays of $B_c^\pm$ mesons} \label{weak}

Motivated by the observation of a large number of exotic $XYZ$ mesons in the decays of the $B^\pm$ and
$B^0$-mesons, as well
as the pentaquark states $P_c(4450)^+$ and $P_c(4380)^+$ in the decays of the $\Lambda_b$-baryons,
\mbox{$\Lambda_b^0 \to (P_c(4380)^+, P_c(4450)^+) K^-$}, with the subsequent decays
 $(P_c(4380)^+, P_c(4450)^+) \to
J/\psi\, p$, we anticipate production of the charged $X_{b0}(5570)^\pm$ and  neutral $X_{b0}(5570)^0$
tetraquark states in weak decays of the $B_c^\pm$ mesons. %
We also emphasize that $B_c^\pm$-decays are a copious, as yet unexplored,
source of hidden $c\bar{c}$ tetraquark states,  via decay modes such as
 $B_c^\pm \to X(3872) \pi^\pm$.  Other candidate tetraquark states in the same family but having different $J^{PC}$ quantum numbers are, likewise, anticipated
 in $B_c^\pm$ decays.

For the weak decays of $B_c^\pm \to B_s^0 \pi^\pm$, $B_c^\pm \to X_{b0}^0 \pi^\pm$,
and $B_c^\pm \to X_{b0}^\pm \pi^0$, the active decay at the
quark level is  $c \to s  u \bar{d}$, with the $\bar{b}$ decay treated as a spectator.
This accounts for approximately 50\% of the $B_c^\pm$ decays~\cite{Gouz:2002kk}.

The effective Hamiltonian for such non-leptonic decays is
\ba
{\cal H}_{\mathrm{eff}}&=& \frac{4G_F}{\sqrt{2}} V_{cs} V_{ud}^* \left[C^{(-)} {\cal O}^{(-)}+ C^{(+)}{\cal O}^{(+)} \label{Hweak1} \right]\\
{\cal O}^{(\pm)}&=& [{\bar{s}}^\alpha \gamma_\mu P_L c_\alpha] [{\bar{u}}^\beta \gamma^\mu P_L d_\beta] \pm  [ \bar{s}^\alpha \gamma_\mu P_L d_\alpha] \nonumber 
\ea
where $G_F$ is the Fermi coupling constant, $V_{ij}$ are the CKM matrix elements,
$\alpha$ and $\beta$ are the color indices, $P_L=\frac{1}{2}(1-\gamma_5)$ and $C^{(\pm)}=(C_1\pm C_2)/2$, $C_{1,2}(\mu)$ being the Wilson coefficients at scale $\mu=m_c,m_b$. We have dropped QCD penguin contributions and $C^{(\pm)}$ are QCD renormalization factors~\cite{glam} computed at a momentum scale equal to the $b$-quark mass, with~\cite{buras1}
\be 
2\, C^{(-)} \simeq 1.4 \quad \quad 2\, C^{(+)} \simeq 0.85\label{buras}
\ee

The amplitude for $B_c^+ \to B_s \pi^+$ can be written in the factorized form, see Fig.~\ref{Feyn}(a)

\begin{equation}
{\cal M}(B_c^+ \to B_s^0 \pi^+)= \frac{G_F}{\sqrt{2}} V_{cs}V_{ud}^* \, (C^{(-)}+C^{(+)})\, \tilde{M} \label{me1} 
\end{equation}
with $(C^{(-)}+C^{(+)})=C_1$
\ba
\tilde{M}&=& \frac{f_{\pi}}{m_\pi^2} q^\mu \langle B_s | \bar{s}\,\gamma_\mu P_L\, c |B_c^+\rangle \nonumber\\
&=&  \frac{f_{\pi}}{m_\pi^2} [ f_{+}(m_\pi^2) (m_{B_c}^2 - m_{B_s}^2) +  f_{-}(m_{\pi}^2)m_\pi^2]
\ea
Here, $f_{\pm}(q^2)$ are the vector current form factors, evaluated at $q^2=m_\pi^2$, which have been
studied  in a number of models (see, for example~\cite{Wang:2008xt} for a comparative evaluation),
 $f_\pi$ is the pion decay constant, $f_\pi =140$~MeV~\cite{pdg}, 
 $C_1$ is the (QCD renormalized) effective Wilson coefficient, estimated
 to be $C_1 \approx 1.1$,  and
the second term above can be neglected, as it is multiplied by $m_\pi^2$.
 With this, the decay width can be evaluated straightforwardly
\begin{equation}
\Gamma (B_c^\pm \to B_s^0 \pi^\pm)= 
 |{\cal M} |^2 \frac{|\bm p_\pi |}{8\pi m_{B_c}^2}\label{decaywd1}
\end{equation}
where $|\bm p_\pi|$ is the $\pi^\pm$ 3-momentum in the rest frame of $B_c^\pm$-meson.

 The branching ratio for 
$B_c^\pm \to B_s^0 \pi^\pm$ has been measured by LHCb 
%
\be
{\cal B}(B_c^+ \to B_s^0 \pi^+) \frac{P(\bar{b} \to B_c^+)}{P(\bar{b} \to B_s)}
= (2.37^{+0.37}_{-0.35}) \times 10^{-3}
\ee
%
Here, $P(\bar{b} \to B_s)$ and $P(\bar{b} \to B_c^+)$ are the fragmentation probabilities. The
ratio of the two probabilities, {\it i.e.}, the ratio of the production rates of $B_c^+$ mesons and
$B_s$ mesons in a $b$-quark jet is estimated to be about $0.02$, yielding a 10\% branching ratio for $B_c^+ \to B_s \pi^+$.
This is the largest branching ratio of any $B$-meson observed in a single channel. 


The decay $B_c^\pm \to X_{b0}^{I=0} (5770) \pi^\pm$ is expected to have a large
branching ratio, as this decay  amplitude, like $B_c^\pm \to B_s^0 \pi^\pm$, is factorizable (see Fig.~\ref{Feyn}(b)).
The relevant matrix element  can be  written down
in an analogous way to that of $B_c^\pm \to B_s^0 \pi^\pm$.
 One now needs to know the hadronic matrix element
$ \langle X_{b0}^{I=0} | \bar{s}\, \gamma_\mu P_L\, c |B_c^+\rangle$.  Recalling that  $X_{b0}^{I=0}$ has $J^P=0^+$,
the transition goes via the axial-vector part of the charged current, yielding an
expression similar to the one for ${\cal M}(B_c^\pm \to B_s^0 \pi^\pm)$ obtained above. However, in this
case, the corresponding hadronic quantity, which we denote by 
 $f_{+}(m_\pi^2)^{B_c X_{b0}}$, is unknown.  This can be
calculated using QCD sum rules or lattice QCD, as it involves the axial-current matrix element of a
single hadron $\to$ single hadron transition.  
 In the diquark model at hand, it is expected to be not
too different from $f_{+}(m_\pi^2)^{B_c B_s^0}$, as the heavy flavor content of the $X_{b0}^0$ and $B_s^0$ is
the same, namely $\bar{b} s$. 

 Denoting the ratio of the two form factors as
 $F(X_{b0}/B_s)\equiv f_{+}(m_\pi^2)^{B_c X_{b0}}/f_{+}(m_\pi^2)^{B_c B_s^0}$, the relative
branching ratios can be expressed as
\ba
&&\frac{{\cal B}(B_c^\pm \to X_{b0}^{I=0} \pi^\pm)}
       {{\cal B}(B_c^\pm \to B_s^0 \pi^\pm)}=
\nonumber\\
&&= F(X_{b0}/B_s)^2 \frac{(m_{B_c}^2 - m_{X_{b0}}^2)^2} {(m_{B_c}^2 - m_{B_s}^2)^2}
 \frac{|\bm p_\pi |^{B_c \to X_{b0}\pi}}{|\bm p_\pi|^{B_c \to B_{s} \pi}} \label{relbr}
\ea
With the known masses, and using our estimate $m(X_{b0}^{I=0})=5.770$ GeV, we get a branching ratio of 1(2)\% for the
decay $B_c^\pm \to X_{b0}^{I=0} \pi^\pm$ for an assumed value of $F(X_{b0}/B_s)^2=0.5 (1)$. Given the large sample of
$B_c^\pm$ already available and in forthcoming LHC runs, this branching ratio is measurable in the decay
mode $B_c^\pm \to (B_s^0 \pi^0) \pi^\pm$, assuming a good $\pi^0$  detection efficiency. 

 We expect the corresponding branching ratio for the decay
$B_c^\pm \to X_{b0}^\pm \pi^0 \to (B_s^0 \pi^\pm ) \pi^0$ to be multiplied by a factor $C^{(-)2}/(C^{(-)}+C^{(+)})^2 \simeq 0.62$. In fact, ${\cal O}^{(-)}$ and ${\cal O}^{(+)}$ contribute equally to  $B_c$ decay into $X_{b0}^{I=0}$, Fig.~\ref{Feyn}(b), while only ${\cal O}^{(-)}$ contributes to the decay into $X_{b0}^{I=1}$, due to color antisymmetry of the final $us$ pair ,~Fig.~\ref{Feyn}(c) (this is similar to the Pati and Woo argument~\cite{patiwoo} to derive the $\Delta I=1/2$ rule, {\it i.e.} flavor antisymmetry, in non-leptonic baryon decays).
This pattern could be modified by non-perturbative effects, as also seen in a number of similar  $B^\pm$ and $B^0$ decays \cite{pdg}.

We now discuss the $B_c^+$ decays leading to the bound $c\bar{c}$ tetraquarks. This requires the
quark decay \mbox{$\bar{b} \to \bar{c} u \bar{d}$}, with the $c$-quark in $B_c^+$ acting as an spectator quark.
The benchmark decay for this class is \mbox{$B_c^\pm \to J/\psi\, \pi^\pm$}. Requiring now the excitation of
 a $q\bar{q}$ pair, followed by quark recombination, leads to decays such as
\mbox{$B_c^\pm \to X(3872)^{I=0}\pi^\pm$} and $B_c^\pm \to X(3872)^{I=1\pm,0}~\pi^{0,\pm}$. 
These diagrams allow access to both the $I=0$ (isosinglet) and the $I=1$ (isotriplet) partners of
the  $X(3872)$, decaying, respectively, to $J/\psi\, \omega$ and $J/\psi\, \rho^0$, as well as
the decay of the charged partner $X(3872)^\pm \to J/\psi\, \rho^\pm$, in addition, possibly, to $\bar{D}^* D$ decays.  There would be enough phase space to observe the corresponding $P$-states as well.  

Again, we expect the decay $B_c^\pm \to X(3872)^{(I=0)} \pi^\pm$ to have a large branching ratio,
which is similar to $B_c^\pm \to J/\psi\, \pi^\pm$, as both are 
factorizable processes and are proportional to $C^{(-)}+C^{(+)}$. The decays of $B_c^\pm$ to 
the $[c q][\bar{c} \bar{q^\prime}]$-tetraquarks have the potential to map out a large number
 of anticipated states in this sector. 

\section{Concluding Remarks}
The observation of the $X(5568)$ by D0, with $X(5568) \to B_s^0 \pi^\pm$, having $M=5568$~MeV and $\Gamma=22$~MeV,
 has not been confirmed by LHCb. It remains to be seen if a state with the quark flavors
$b \bar{s} u \bar{d}$ exists in nature, with a different mass, decay pattern, and width.
  In this paper we have used the diquark-antidiquark picture
 to give predictions about the mass spectrum of the lowest $S$-state, $X_{b0}$ and its $J^P=1^+$ partners,
 both in  charm and bottom sectors.
Our estimates set the mass of the lowest such state in the $b$-quark sector at around 5770~MeV,
 somewhat below the $BK$ threshold. Within the errors of our approach, $X_{b0}^+$ could lie
just above this threshold, and one has to look for it in the decay $X_{b0}^+ \to B^+\bar K^0$. However,
  $X_{b0}^+$  may as well reveal itself as
 a resonating $B_s \pi$ state, or  not manifest at all, if below threshold, as discussed in~\cite{fm}.
 
 Here we propose to search tetraquark states in the decays of the $B_c^\pm$ mesons, $B_c^\pm \to X_{b0}^0 \pi^\pm$
and  $B_c^\pm \to X_{b0}^\pm \pi^0$ and have argued that some of these decay modes may have a
large branching ratio. This requires a good $\pi^0$-detection efficiency, which we advocate to
improve in hadron collider experiments, such as the LHCb. The two detached vertices (of
 the $B_c^\pm$ and $B_s^0$) may help in reducing the background.

 So far, only a handful of $B_c^\pm$
decays have been observed~\cite{pdg}, and it is worthwhile to put in a dedicated effort to
increase this database.  
 Apart from the possibility of observing the tetraquark states of the $B_s\pi$ variety,
we anticipate several bound $c\bar{c}$ tetraquark states, which emerge from the decay
 $B_c^+\to (c\bar{c}) u\bar{d}$, followed by a $q\bar{q}$ excitation from the vacuum.
These would lead to decays such as $B_c^\pm \to X(3872)^0 \pi^\pm$ and $B_c^\pm \to X(3872)^\pm \pi^0$,
as well as to other related tetraquark states. They should be searched for at the LHC, and also at
Belle-II, if the $e^+e^-$ center of mass energies could reach the $B_c^+ B_c^-$ threshold.

\vspace{1cm}

\textbf{Acknowledgments:} We thank Ishtiaq Ahmed, Jamil Aslam and Abdur Rahman for correspondence on
the mass spectrum and Tim Gershon and Sheldon Stone for useful discussions on the experimental aspects.



\begin{thebibliography}{99}


\bibitem{D0:2016mwd} 
  V.~M.~Abazov {\it et al.} [D0 Collaboration],
  [\href{http://arxiv.org/abs/1602.07588}{arXiv:1602.07588 [hep-ex]}].
  
\bibitem{others} 
  W.~Chen, H.~X.~Chen, X.~Liu, T.~G.~Steele and S.~L.~Zhu,
  \href{http://arxiv.org/abs/1602.08916}{arXiv:1602.08916 [hep-ph]};
    W.~Wang and R.~Zhu,
  \href{http://arxiv.org/abs/1602.08806}{arXiv:1602.08806 [hep-ph]};
    Z.~G.~Wang,
  \href{http://arxiv.org/abs/1602.08711}{arXiv:1602.08711 [hep-ph]};
    S.~S.~Agaev, K.~Azizi and H.~Sundu,
  \href{http://arxiv.org/abs/1602.08642}{arXiv:1602.08642 [hep-ph]};
    C.~J.~Xiao and D.~Y.~Chen,
  \href{http://arxiv.org/abs/1603.00228}{arXiv:1603.00228 [hep-ph]};
    S.~S.~Agaev, K.~Azizi and H.~Sundu,
  \href{http://arxiv.org/abs/1603.00290}{arXiv:1603.00290 [hep-ph]};
  Y.~R.~Liu, X.~Liu and S.~L.~Zhu,
  \href{http://arxiv.org/abs/1603.01131}{arXiv:1603.01131 [hep-ph]}.
  C.~M.~Zanetti, M.~Nielsen and K.~P.~Khemchandani,
  arXiv:1602.09041 [hep-ph].
  X.~H.~Liu and G.~Li,
  arXiv:1603.00708 [hep-ph].
  X.~G.~He and P.~Ko,
  arXiv:1603.02915 [hep-ph].
  Y.~Jin and S.~Y.~Li,
  arXiv:1603.03250 [hep-ph].
  L.~Tang and C.~F.~Qiao,
  arXiv:1603.04761 [hep-ph].
%
\bibitem{Burns:2016gvy} 
  T.~J.~Burns and E.~S.~Swanson,
  arXiv:1603.04366 [hep-ph].
  \bibitem{tantaloni} 
  A.~Esposito, M.~Papinutto, A.~Pilloni, A.~D.~Polosa and N.~Tantalo,
  Phys.\ Rev.\ D {\bf 88}, no. 5, 054029 (2013)
  [\href{http://arxiv.org/abs/1307.2873}{arXiv:1307.2873 [hep-ph]}];
    A.~L.~Guerrieri, M.~Papinutto, A.~Pilloni, A.~D.~Polosa and N.~Tantalo,
  PoS LATTICE {\bf 2014}, 106 (2015)
  [\href{http://arxiv.org/abs/1411.2247}{arXiv:1411.2247 [hep-lat]}].

%
\bibitem{LHCb:2016ppf} 
  The LHCb Collaboration [LHCb Collaboration],
  ``Search for structure in the $B_s^0\pi^\pm$ invariant mass spectrum,''
  LHCb-CONF-2016-004, CERN-LHCb-CONF-2016-004.

\bibitem{tetra0}L.~Maiani, F.~Piccinini, A.~D.~Polosa and V.~Riquer,
  Phys.\ Rev.\ Lett.\  {\bf 93} (2004) 212002

%
\bibitem{tetra1} 
  L.~Maiani, F.~Piccinini, A.~D.~Polosa and V.~Riquer,
  Phys.\ Rev.\ D {\bf 71}, 014028 (2005)
  [\href{http://arxiv.org/abs/hep-ph/0412098}{hep-ph/0412098}].
\bibitem{tetra2}
    L.~Maiani, F.~Piccinini, A.~D.~Polosa and V.~Riquer,
  Phys.\ Rev.\ D {\bf 89}, 114010 (2014)
  [\href{http://arxiv.org/abs/1405.1551}{arXiv:1405.1551 [hep-ph]}].
  %
  \bibitem{tetrab}  A.~Ali, C.~Hambrock and W.~Wang,
  Phys.\ Rev.\ D {\bf 85} (2012) 054011.
  \bibitem{tetra3}
    A.~Ali, L.~Maiani, A.~D.~Polosa and V.~Riquer,
  Phys.\ Rev.\ D {\bf 91}, 1, 017502 (2015)
  [\href{http://arxiv.org/abs/1412.2049}{arXiv:1412.2049 [hep-ph]}].
%
 
 \bibitem{lhcbat} 
  R.~Aaij {\it et al.} [LHCb Collaboration],
  Phys.\ Rev.\ Lett.\  {\bf 110}, no. 15, 151803 (2013)
  doi:10.1103/PhysRevLett.110.151803
  [arXiv:1211.5994 [hep-ex]].

  
\bibitem{fm} 
A.~Esposito, A.~Pilloni and A.~D.~Polosa,
  arXiv:1603.07667 [hep-ph].

\bibitem{Aaij:2013cda} 
  R.~Aaij {\it et al.} [LHCb Collaboration],
  Phys.\ Rev.\ Lett.\  {\bf 111}, no. 18, 181801 (2013)
  doi:10.1103/PhysRevLett.111.181801
  [arXiv:1308.4544 [hep-ex]].

%
 \bibitem{jaffe}
  R.~L.~Jaffe,
  Phys.\ Rept.\  {\bf 409}, 1 (2005)
  [\href{http://arxiv.org/abs/hep-ph/0409065}{hep-ph/0409065}];
  L.~Maiani, F.~Piccinini, A.~D.~Polosa and V.~Riquer,
  Phys.\ Rev.\ Lett.\  {\bf 93}, 212002 (2004)
  [\href{http://arxiv.org/abs/hep-ph/0407017}{hep-ph/0407017}];
  C.~Alexandrou, P.~de Forcrand and B.~Lucini,
  Phys.\ Rev.\ Lett.\  {\bf 97}, 222002 (2006)
  [\href{http://arxiv.org/abs/hep-lat/0609004}{hep-lat/0609004}].
  


\bibitem{lepton} 
  S.~Stone and J.~Urheim,
  eConf C {\bf 030603}, MAR05 (2003)
  [AIP Conf.\ Proc.\  {\bf 687}, 96 (2003)]
  [\href{http://arxiv.org/abs/hep-ph/0308166}{hep-ph/0308166}]; B.~Aubert {\it et al.} [BaBar Collaboration],
  Phys.\ Rev.\ D {\bf 74} (2006) 032007
  [hep-ex/0604030];
  B.~Aubert {\it et al.} [BaBar Collaboration],
  Phys.\ Rev.\ D {\bf 80}, 092003 (2009)
  [\href{http://arxiv.org/abs/0908.0806}{arXiv:0908.0806 [hep-ex]}].

\bibitem{Gouz:2002kk} 
  I.~P.~Gouz, V.~V.~Kiselev, A.~K.~Likhoded, V.~I.~Romanovsky and O.~P.~Yushchenko,
  Phys.\ Atom.\ Nucl.\  {\bf 67}, 1559 (2004)
  [Yad.\ Fiz.\  {\bf 67}, 1581 (2004)]
  doi:10.1134/1.1788046
  [hep-ph/0211432].

\bibitem{glam} M.~K.~Gaillard, B.~W.~Lee,  Phys.\ Rev.\ Lett.\  {\bf 33}, 108 (1974); G.~Altarelli, L.~Maiani, Phys.\ Lett.\  {\bf B52}, 351.
\bibitem{buras1} see e.g.: A.~J.~Buras, {\it Weak Hamiltonian, CP violation and rare decays}, hep-ph/9806471.



\bibitem{Wang:2008xt} 
  W.~Wang, Y.~L.~Shen and C.~D.~Lu,
  Phys.\ Rev.\ D {\bf 79}, 054012 (2009)
  doi:10.1103/PhysRevD.79.054012
  [arXiv:0811.3748 [hep-ph]].

\bibitem{patiwoo}  J.~C.~Pati and C.~H.~Woo,
  Phys.\ Rev.\ D {\bf 3} (1971) 2920.

\bibitem{pdg}
  K.~A.~Olive {\it et al.} [Particle Data Group Collaboration],
  Chin.\ Phys.\ C {\bf 38}, 090001 (2014).


  
  
\end{thebibliography}
\end{document}